\documentclass[floatfix,aps,amsmath,nofootinbib,onecolumn,10pt]{revtex4}
\pdfoutput=1

\usepackage{graphicx}
\usepackage{bm}

\def\half{{1\over 2}}

\def\half{{1\over 2}}
\def\({\left(}
\def\){\right)}
\def\[{\left[}
\def\]{\right]}

\def\e{\begin{equation}}
\def\q{\end{equation}}
\def\m{\begin{eqnarray}}
\def\n{\end{eqnarray}}

\begin{document}

\title{Constraint on the abundance of primordial black holes in dark matter from Planck data}

\author{Lu Chen\footnote{chenlu@itp.ac.cn}, Qing-Guo Huang\footnote{huangqg@itp.ac.cn} and Ke Wang \footnote{wangke@itp.ac.cn}}
\affiliation{$^1$CAS Key Laboratory of Theoretical Physics,\\ Institute of Theoretical Physics, \\Chinese Academy of Sciences, Beijing 100190, China\\
and\\
$^2$School of Physical Sciences, \\University of Chinese Academy of Sciences,\\ No. 19A Yuquan Road, Beijing 100049, China}

\date{\today}

\begin{abstract}

We use Planck data released in 2015 to constrain the abundance of primordial black holes (PBHs) in dark matter in two different reionization models (one is the instantaneous reionization and the other is the asymmetric reionization), and significantly improve the existing upper limits on the abundance of PBHs by around two orders of magnitude. These new limits imply that the event rates of mergers of PBH binaries (Gpc$^{-3}$ yr$^{-1}$) are less than $0.002$ for $M_\text{pbh}=30M_\odot$, $5$ for $M_\text{pbh}=10M_\odot$ and $2000$ for $M_\text{pbh}=2M_\odot$ at $95\%$ confidence level (C.L.), and thus the gravitational-wave event GW150914 is very unlikely produced by the merger of a PBH binary.

\end{abstract}

\pacs{???}

\maketitle


\section{Introduction}

In the Big Bang Model there were large inhomogeneities and the density of overdensed regions kept growing and finally collapsed to Primordial Black Holes (PBHs) \cite{Hawking:1971ei,Carr:1974nx}.
They accrete matter around them and emit radiations at the same time.
PBHs less than $10^{15}$ g would evaporate away completely within the age of the universe because the Hawking radiation overweighted the accretion. Only the large enough ones would be stable and grow larger by accretion or merging with each other, persisting to present. Substantial massive non-evaporationg PHBs may produce some detectable effects on the universe, such as the star formation, the constitution of dark matter, the anisotropies and spectral distortions of the cosmic microwave background (CMB) and so on.

In principle the massive PBHs are non-relativistic and almost collisionless, and hence there is nothing to prevent that the massive PBHs make up the cold dark matter (CDM), at least, partly. Despite the weakly interacting massive particles (WIMPs) are the most acknowledged candidates for CDM, up to now, there is no evidence from all kinds of experiments, such as the Large Hadron Collider (LHC) \cite{ATLAS:2016hao}, the Large Underground Xenon (LUX) dark matter experiment \cite{Akerib:2015rjg}, Fermi LAT (Fermi Large Area Telescope) \cite{FermiLAT:2011ab}, the Alpha Magnetic Spectrometer (AMS-02) \cite{Accardo:2014lma} and so on. It urges us to turn our attention to other possible candidates, such as PBHs. On the other hand, this year LIGO (Laser Interferometer Gravitationsl-Wave Observatory) claimed the discovery of the gravitational waves from two merging black holes \cite{Abbott:2016blz,Abbott:2016nmj}. It has rekindled the theories that PBHs are the component of dark matter \cite{Bird:2016dcv,Sasaki:2016jop,Kashlinsky:2016sdv,Cholis:2016kqi}.

Up to now there is no evidence for the existence of PBHs.
The various constraints on the abundance of PBHs in dark matter, namely
\m
f_\text{pbh} = \frac{\Omega _\text{pbh}}{\Omega _\text{CDM}}
\n
has been given in the literature. The PBHs with mass around $10^{15}$ g would be evaporating today and the extragalactic $\gamma$-rays provides a stringent constraint on them \cite{Carr:2009jm}, i.e. $f_\text{pbh}\lesssim 2\times 10^{-8} (M_\text{pbh}/M_*)^{3.2}$ for $M_\text{pbh}>M_*=5\times 10^{14}$ g. Since the neutron star gets destroyed in a very short time due to the accretion onto PBH once a PBH is captured by it, this effect gives a constraint on the abundance of PBHs ($f_\text{pbh}<0.05$ for $3\times 10^{18} \text{g} \lesssim M_\text{pbh}\lesssim 10^{24}\text{g}$) \cite{Capela:2013yf}. The observations of stars in the Magellanic Clouds for microlensing events caused by MACHO (massive astrophysical compact halo object) are also used to test the hypothesis that MACHO could be a major component of the dark matter halo of the Milky Way galaxy, and,unfortunately, both EROS-2 (Eath Resources Observation Satellite) and OGLE (Optical Gravitational Lensing Experiment) rule out MACHOs as the majority of Galactic dark matter over the range $0.6\times 10^{-7} M_\odot<M_\text{pbh}<15 M_\odot$: $f_\text{pbh}<0.04$ for $10^{-3} M_\odot \lesssim M_\text{pbh}\lesssim 10^{-1}M_\odot$ and $f_\text{pbh}<0.1$ for $10^{-6} M_\odot \lesssim M_\text{pbh}\lesssim M_\odot$ in \cite{Tisserand:2006zx};  $f_\text{pbh}\lesssim 0.06$ for $0.1 M_\odot \lesssim M_\text{pbh}\lesssim 0.4 M_\odot$ and $f_\text{pbh}\lesssim 0.2$ for $0.4 M_\odot \lesssim M_\text{pbh}\lesssim 15 M_\odot$ in \cite{Wyrzykowski:2010mh,Wyrzykowski:2011tr}.
PBHs with large mass could have a large luminosity at early times due to accretion and then could have an important effect on the thermal history of the Universe even if their density is small \cite{Carr:1981}. Modeling the accumulation of dark matter around PBHs and the proper motion of PBHs, Ricotti et al. figured out the effects on the CMB from the gas accretion onto PBHs and obtained upper limits on the abundance of PBHs with mass larger than $0.1 M_\odot$ in \cite{Ricotti:2007au}.
See \cite{Carr:2016drx} for a more comprehensive summary about the constraints on the abundance of PBHs.

In this paper we will constrain the abundance of PBHs in dark matter by adopting the Planck data released in 2015 \cite{Ade:2015xua} and two different models for the reionization (one is the so-called instantaneous reionization and the other is the asymmetric reionization proposed in \cite{Douspis:2015nca}).

This paper is organized as follows. In Sec.~\ref{cmb}, we sketch out the effects of PBHs on the CMB temperature anisotropies and polarizations. In Sec.~\ref{abundance}, we utilize Planck data to constrain the abundance of PBHs in dark matter. In Sec.~\ref{er}, we apply the results obtained in Sec.~\ref{abundance} to estimate the upper limit on the event rate of mergers of PBH binaries. A summery and discussion are given in Sec.~\ref{sd}.

\section{Effects of primordial black holes on the CMB}
\label{cmb}

The PBHs with a wide range of masses could have formed before any other astrophysical objects has been formed. For example, the direct gravitational collapse of an order of unity primordial density inhomogeneity can lead to formation of PBHs when such a perturbation mode re-enters the horizon \cite{Hawking:1971ei,Carr:1974nx} during radiation-dominated era. The mass of the PBH is approximately equal to the horizon mass at the time of formation, namely
\m
M_\text{pbh}\sim {c^3 t\over G}\sim 10^5  \({t\over 1\ \text{s}}\) M_\odot.
\n
The small mass PBHs can have been formed in the early universe. By contrast, black holes forming at the present epoch could never be smaller than about 1 $M_\odot$. Because the temperature of black hole has temperature
\m
T_\text{bh}={\hbar c^3\over 8\pi G M k_B}
\n
due to its quantum properties \cite{Hawking:1974rv,Hawking:1974sw}, such a black hole is supposed to evaporate completely within the time scale
\m
T_\text{eva}(M_\text{bh})\sim {G^2M_\text{bh}^3\over \hbar c^4}\sim 10^{64} \({M_\text{bh}\over M_\odot}\)^3\ \text{yr}.
\n
Therefore the PBHs with mass smaller than $10^{-18} M_\odot$ would have evaporated by now. However the PBHs with mass larger than $10^{-18} M_\odot$ would still survive today and might be detected. Generally it is considered unlikely that the PBHs form after $1$ s, when $M_\text{pbh}>10^5 M_\odot$, because the physics in this epoch is nicely understood and their formation would affect, for example, primordial nucleosynthesis.

Here we focus on the effects of non-evaporating PBHs on the CMB in \cite{Ricotti:2007au}.
Gas accretion onto the non-evaporating PBHs produces radiations, such as X-rays and UV radiation. The X-rays have a mean free path larger than the mean distance between PBHs, and ionize the gas and heat the intergalactic medium (IGM). The modification of the ionization and thermal history of the universe from the accretion affects the spectrum and anisotropies of the CMB.

The accretion luminosity being responsible for altering the evolution of the IGM can be estimated from the gas accretion rate.
For simplicity, the accretion rate for a point mass $M$ traveling through an hydrogen gas with constant number density $n_\text{gas}$ and sound speed $c_s$ at velocity $v$ with respect to the gas is given by
\m
{\dot M}_b=\lambda 4\pi m_H n_\text{gas}v_\text{eff} r_B^2,
\n
where the mean cosmic gas density is
\m
n_\text{gas}\simeq 200 \({1+z\over 1000}\)^3\ \text{cm}^{-3},
\n
\m
v_\text{eff}\equiv \sqrt{v^2+c_s^2},
\n
and
\m
r_B\equiv {GM\over v_\text{eff}^2}
\n
is the Bondi radius \cite{Bondi:1944jm,Bondi:1952ni}.
Supposing a non-viscous fluid, $\lambda$ is of order unity for spherical accretion onto a point mass. The eigenvalue $\lambda$ for PBHs was calculated in \cite{Ricotti:2007au} where all the effects of the growth of a dark halo around PBHs, the Hubble expansion and the coupling of the CMB radiation to the gas through Compton scattering are considered.
On the other hand, the hydrogen around each PBH is fully ionized by the UV radiation and forms the HII region which may reduce the gas accretion rate if the radius of the HII region is larger than the Bondi radius (i.e. $r_{\text{HII}}>r_B$) and the gas temperature inside the HII region is higher than the temperature outside. Due to the feedback of UV radiation, the accretion luminosity is suppressed by a factor
\m
f_\text{duty}={1\over 1+(r_{\text{HII}}/r_B)^{1/3}}.
\n
Actually the local feedback due to the formation of HII region around PBH is negligible in most cases, and only needed to be taken into account for the massive PBHs with $M_\text{pbh}>10^2\sim 10^3 M_\odot$.

Simulating the ionization, chemical and thermal history of the universe after recombination with the PBHs as the only ionizing source, the authors found that the ionization fraction increases approximately as $x_e\equiv n_e/n_H\propto (1+z)^{-1}$ from $x_e\sim 10^{-3}$ at $z\sim 900$ to values close to $x_e\sim 10^{-1}-10^{-2}$ at $z\sim 10$ in \cite{Ricotti:2007au}.
Therefore, the existence of PBHs can leave an imprint on the CMB power spectra through modifying the ionization fraction.

\begin{figure}[]
\begin{center}
\includegraphics[width=5.5in]{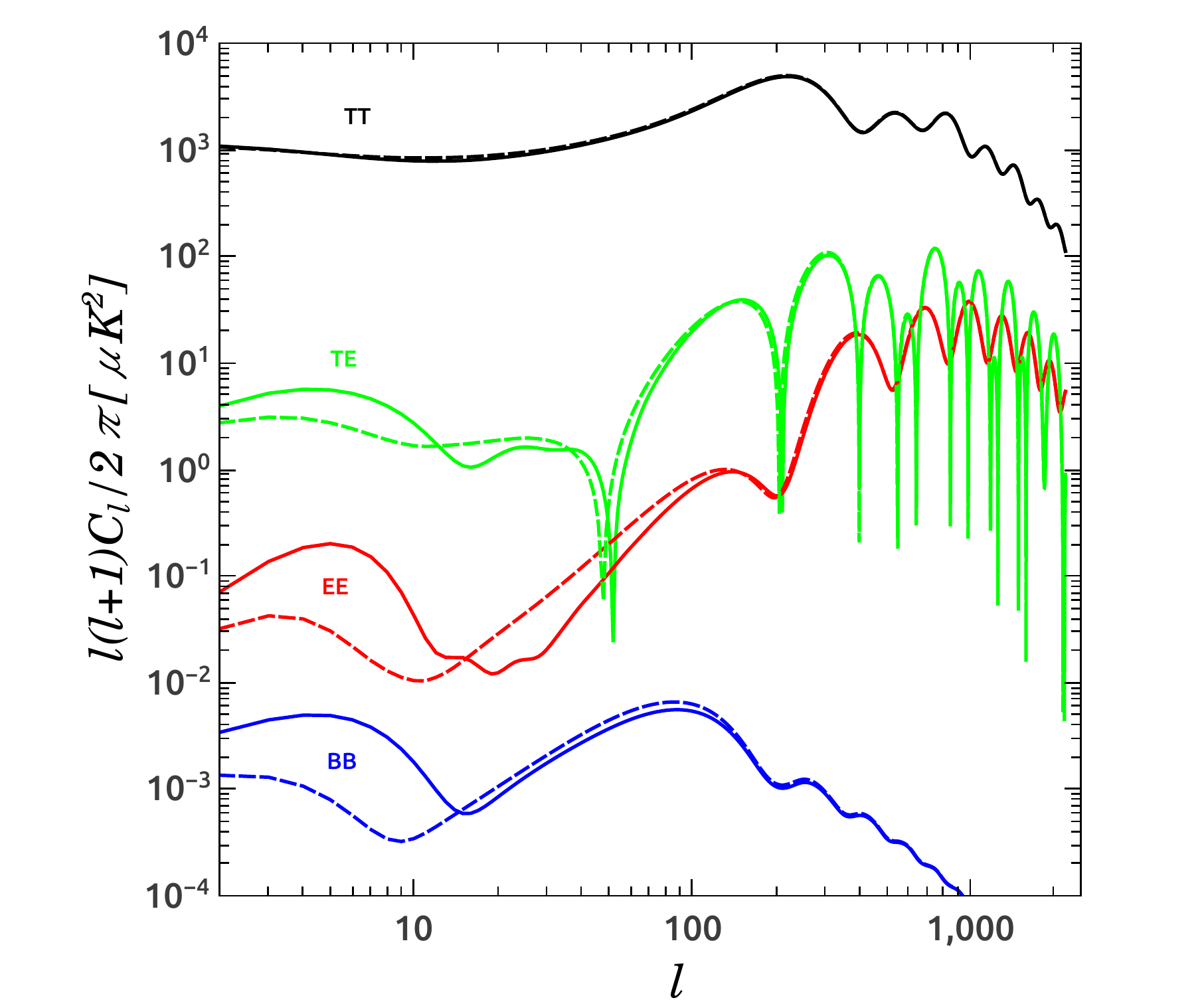} 
\end{center}
\caption{CMB angular power spectra. Here the total optical depth $\tau_e=0.134$ is kept fixed and the tensor-to-scalar ratio is $r=0.1$. The solid curves and dashed curves correspond to $x_{e0}=0$ (without PBHs) and $x_{e0}=1\times 10^{-3}$ respectively. }
\label{fig:cl}
\end{figure}

For the instantaneous reionization model, we parameterize the ionization fraction for including the contribution from PBHs as follows 
\begin{align}\label{xes}
x_e(z)=
\begin{cases}
x_\text{e,rec}(z)+x_\text{e,pbh}(z), ~~\text{for}\ z\geq z_{\textrm{beg}}\ ;\\
{f-x_\text{e,pbh}(z_\text{beg})-x_\text{e,rec}(z_\text{beg})\over 2}\[1+\tanh\(\frac{y(z_{re})-y(z)}{\Delta _y}\) \]+x_\text{e,pbh}(z_\text{beg})+x_\text{e,rec}(z_\text{beg}), ~~\text{for}\ z<z_{\textrm{beg}}\ .
\end{cases}
\end{align}
where $f=1+n_{He}/n_H=1.08$ is the ionization fraction of a fully ionized universe, the $x_\text{e,rec}$ is the ionization fraction of recombination history given by RECFAST \cite{Seager:1999bc}, and
\m
x_\text{e,pbh}(z)=\text{min}\[x_{e0}\(1+z\over 10^3\)^{-1},0.1\]
\n
is the ionization fraction produced by PBHs. Both the redshift of a half ionized universe $z_\text{re}$ and the ionization fraction produced by PBHs $x_{e0}$ at $z=1000$ are taken as free parameters. Here the width  $\Delta_z$ and the redshift at the beginning of reionization by other sources $z_\text{beg}$ are set as those in the CAMB code, namely 
\m
\Delta_z&=&0.5, \\
z_\text{beg}&=&z_\text{re}+8\times\Delta_z, \\
y(z)&=&(1+z)^{\frac{3}{2}}, \\
\Delta _y&=&1.5\sqrt{1+z_\text{re}} \Delta _z.
\n
Roughly speaking, the effects on the CMB power spectra from PBHs and other ionizing sources are uncorrelated, and hence the total optical depth can be separated into two parts, namely 
\m
\tau_e=\tau_\text{e,rei}+\Delta \tau_e,
\n
where 
\m
\tau_e\equiv \int_0^{z_*} x_e(z) n_H(z) \sigma_T {dz\over H(1+z)}
\n
is the total optical depth,  
\m
\Delta\tau_e= \int_{z_\text{beg}}^{z_*} x_\text{e,pbh}(z) n_H(z) \sigma_T {dz\over H(1+z)},
\n
is the optical depth contributed by PBHs, $z_*$ is the redshift of recombination and $\sigma_T$ is the Thomson scattering cross-section. 
The effects on the CMB angular power spectra from PBHs are illustrated in Fig.~\ref{fig:cl} where the total optical depth $\tau_e=0.134$ is kept fixed. 
Since the temperature power spectra at all scales has been formed at the time of recombination, the whole spectra are  suppressed by a factor of $e^{-2\tau_e}$. 
However, PBHs effectively provide an ``early-time reionization" (earlier than the time of instantaneous reionization by other ionizing sources), and then the polarization spectra are enhanced by $\Delta\tau_e^2$ in the range $10\lesssim \ell \lesssim 100$. Because the total optical depth is kept fixed, $\tau_\text{e,rei}<\tau_e=0.134$ and then the CMB polarization spectra around the reionization peak (fully ionized by other ionizing sourced at low redshift) become smaller compared to the case without PBHs.


In \cite{Douspis:2015nca}, the authors proposed a new data-motivated parametrization of the history of the average ionization fraction which is asymmetric in the redshift. Here we also take this parameterization into account. Similar to the former case, we modify the ionization fraction to be 
\begin{align}\label{xea}
x_e(z)=
\begin{cases}
x_\text{e,rec}(z)+x_\text{e,pbh}(z), ~~z\geq z_{\textrm{beg}}\ ;\\
f\times Q_\text{HII}(z), ~~z<z_{\textrm{beg}}\ ,
\end{cases}
\end{align}
where the redshift at the beginning of reionization satisfies $x_\text{e,rec}(z_{\textrm{beg}})+x_\text{e,pbh}(z_{\textrm{beg}})=1.08\times Q_\text{HII}(z_{\textrm{beg}})$, and the volume filling factor of HII evolves as
\begin{align}\label{}
Q_\text{HII}(z)=
\begin{cases}
Q_\text{HII}(z_p)e^{-\lambda(z-z_p)}, ~~z_p\leq z<z_{\textrm{beg}}\ ;\\
1-(1-Q_\text{HII}(z_p))\(1+z\over 1+z_p\)^3, ~~z<z_p\ .\\
\end{cases}
\end{align}
Following \cite{Douspis:2015nca,Oldengott:2016yjc}, we also fix the pivot redshift $z_p=6.1$ and $Q_\text{HII}(z_p)=0.99986$ and take the evolution rate $\lambda$ as a free parameter.


\section{Constraint on the abundance of primordial black holes from Planck data}
\label{abundance}

In this section we use the data combination of Planck TT,TE,EE+lowP+lensing data released in 2015 to constrain the abundance of PBHs in dark matter in two different reionization models separately. Note that the ionization fraction $x_e(z)$ in CAMB is modified to be those given in Eqs.~\eqref{xes} and \eqref{xea} for the instantaneous reionization model and the asymmetric reionization model respectively. In summarize, the free parameters needed to be fitted in these two models are \{$\omega_b$,$\omega_c$,$100\theta_{\textrm{MC}}$,$n_s$,$\textrm{ln}(10^{10}A_s)$,$z_\text{re}$,$x_{e0}$\} and \{$\omega_b$,$\omega_c$,$100\theta_{\textrm{MC}}$,$n_s$,$\textrm{ln}(10^{10}A_s)$,$\lambda$,$x_{e0}$\} respectively. Here $\omega_b$ is the physical density of baryons today, $\omega_c$ is the physical density of cold dark matter today, $\theta_{\textrm{MC}}$ is the ratio between the sound horizon and the angular diameter distance at the decoupling epoch, $n_s$ is the scalar spectral index and $A_s$ is the amplitude of the power spectrum of primordial curvature perturbations at the pivot scale $k_p=0.05$ Mpc$^{-1}$.

\begin{figure}[]
\begin{center}
\includegraphics[width=4in]{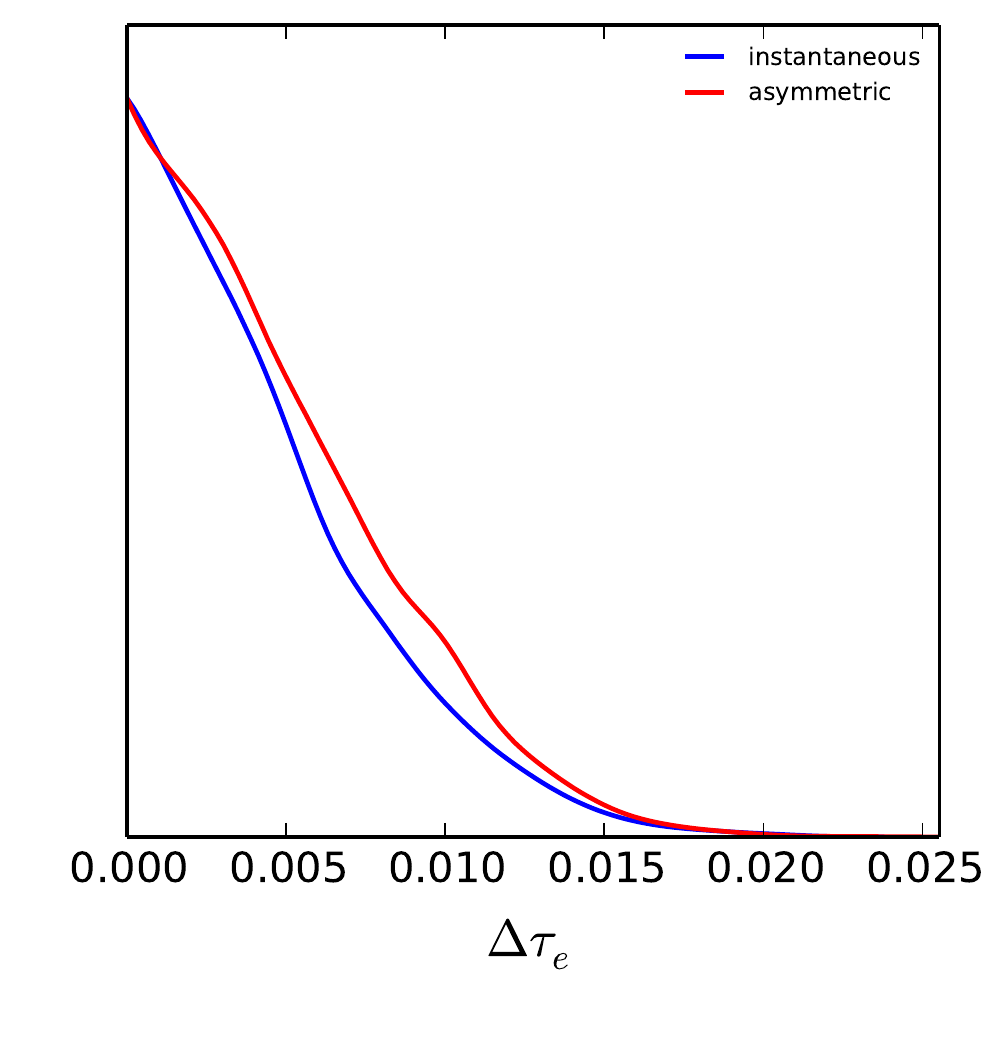}
\end{center}
\caption{The likelihood of $\Delta\tau_e$ in the instantaneous and asymmetric reionzation models. }
\label{fig:taupbh}
\end{figure}

\begin{figure}[]
\begin{center}
\includegraphics[width=4in]{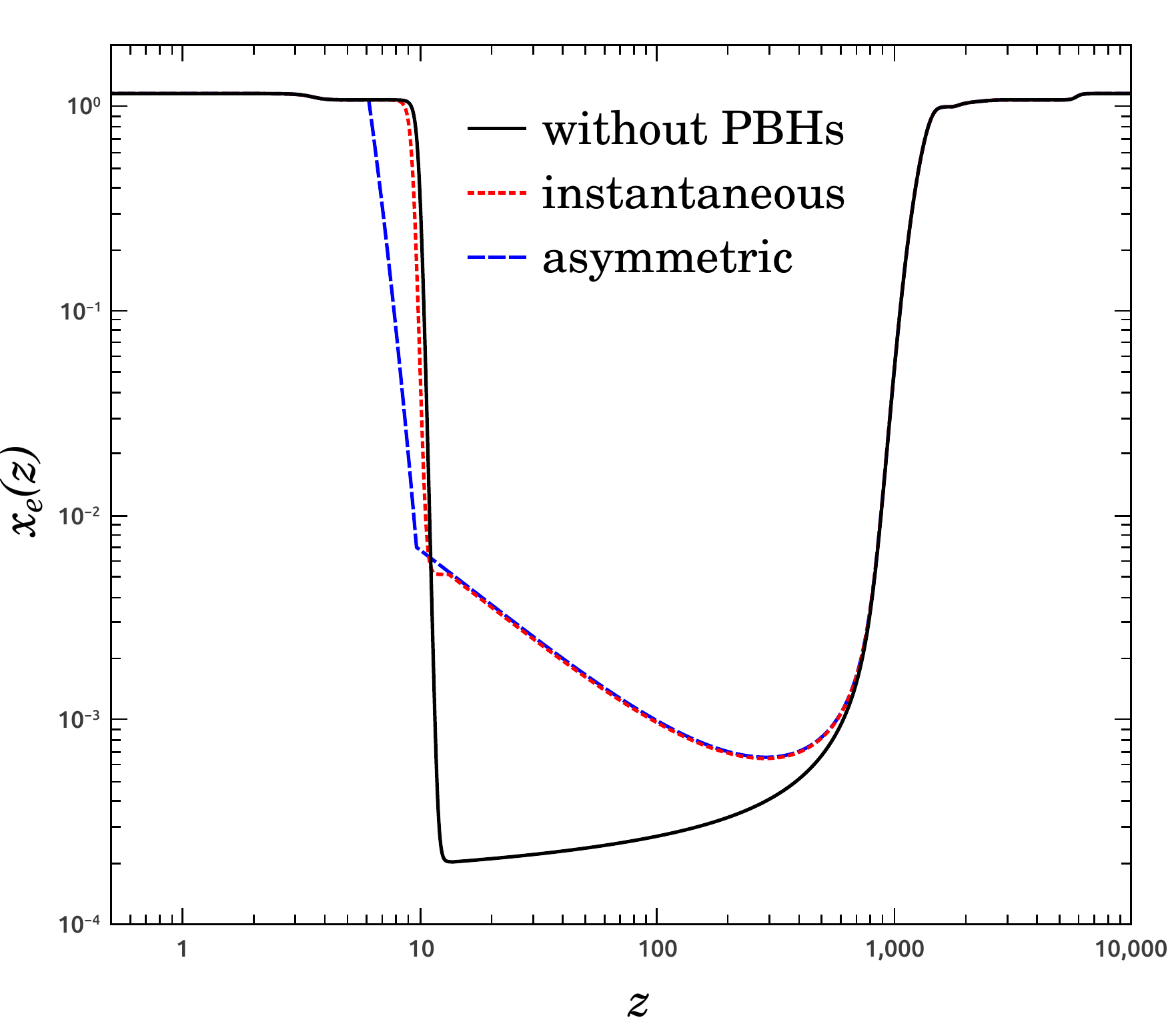}
\end{center}
\caption{The evolution of ionization fraction $x_e(z)$ for different reionization models. The black solid curve corresponds to the instantaneous reionization model without PBHs, and the red dotted and blue dashed curves illustrate the $95\%$ limits for the instantaneous and asymmetric reionization models with PBHs, respectively. }
\label{fig:xe}
\end{figure}

To constrain the cosmological parameters, we refer to the Markov Chain Monte Carlo sampler (CosmoMC) \cite{Lewis:2002ah}. For the instantaneous reionization model, we find 
\m
x_{e0}&<&3.3\times10^{-5}\ (68\%\ \text{C.L.}),\\
x_{e0}&<&7.0\times10^{-5}\ (95\%\ \text{C.L.}),
\n
and then 
\m
\Delta \tau_e&<&0.005\ (68\%\ \text{C.L.}),\\
\Delta \tau_e&<&0.012\ (95\%\ \text{C.L.}). 
\n
The redshift of reionization reads 
\m
z_\text{re}=7.9_{-1.3}^{+1.5}\ (68\%\ \text{C.L.}). 
\n
For the asymmetric reionization model, our results are  
\m
x_{e0}&<&3.7\times10^{-5}\ (68\%\ \text{C.L.}),\\
x_{e0}&<&7.2\times10^{-5}\ (95\%\ \text{C.L.}),
\n 
and then 
\m
\Delta \tau_e&<&0.006\ (68\%\ \text{C.L.}),\\
\Delta \tau_e&<&0.012\ (95\%\ \text{C.L.}). 
\n
In this model, the evolution rate is $\lambda=1.2^{+0.4}_{-0.9}$ at $68\%$ C.L.. These two reionization models give almost the same results. The likelihoods of $\Delta\tau_e$ in these two models show up in Fig.~\ref{fig:taupbh}. 
In addition, the $95\%$ limits of ionization fraction for both the instantaneous and asymmetric reionization models with PBHs are illustrated in Fig.~\ref{fig:xe}. 

\begin{figure}[]
\begin{center}
\includegraphics[width=5in]{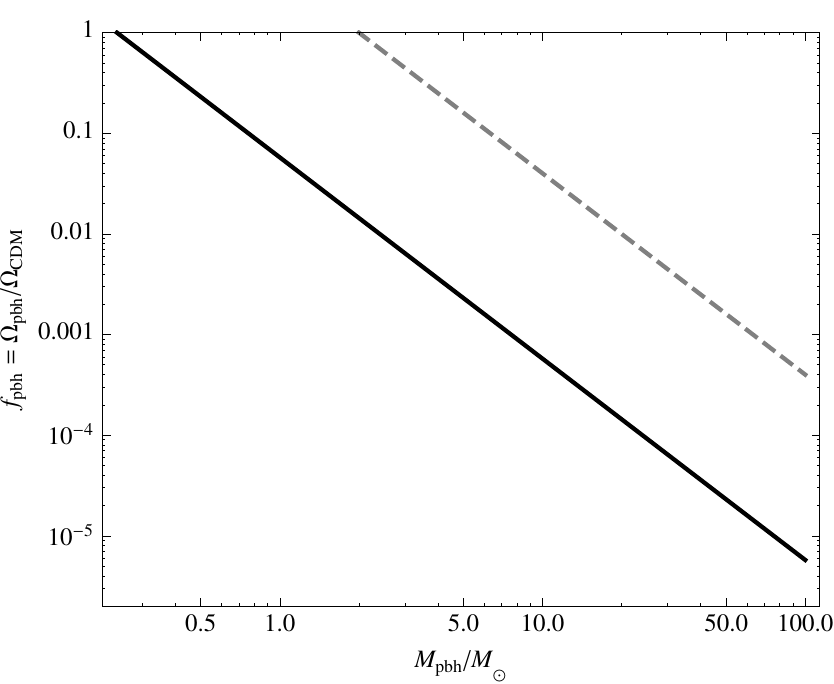}
\end{center}
\caption{The upper limit on the abundance of PBHs in dark matter at $95\%$ confidence level. Here the black solid line represents our results and the grey dashed line corresponds to the constraints from WMAP 3-year data.}
\label{fig:abundance}
\end{figure}

According to the simulation, in \cite{Ricotti:2007au}, the authors found that the value of $\Delta\tau_e$ can be parametrized by a function of mass and abundance of PBHs in dark matter as follows
\m
\Delta\tau_e=0.05\({M_\text{pbh} \over M_\odot}\)f_\text{pbh}^{\half} 
\label{taupbh}
\n
for $M_\text{pbh} < 10^2 M_{\odot}$. Non-zero value of $\Delta\tau_e$ may imply the existence of PBHs. Unfortunately, we only obtain the upper limits on $\Delta\tau_e$ by using the Planck data. In \cite{Ricotti:2007au}, the upper limit on $\Delta\tau_e$ is $0.1$ at $95\%$ C.L. from WMAP 3-year data. Because $f_\text{pbh}\sim (\Delta\tau_e)^2$ and the upper limit on $\Delta\tau_e$ obtained in this paper is roughly one order of magnitude smaller than that in \cite{Ricotti:2007au}, the upper limits on the abundance of PBHs in dark matter is improved by around two orders of magnitude. See the upper limits on the abundance of PBHs in Fig.~\ref{fig:abundance}. \footnote{Because there are typos in Eq.~(42) of \cite{Ricotti:2007au}, one can get the limit on the abundance of PBHs with large mass by simply rescaling Fig.~9 in \cite{Ricotti:2007au}. }

In addition, PBHs have high probability of forming binaries, and the accretion geometry for the binaries becomes disk-like \cite{Ricotti:2007au,Hayasaki:2009ug} once the orbital velocities of the PBHs are larger than the effective translational velocity of the system. Usually the formation of a disk increases the radiative efficiency compared to the spherical case and has more power to reionize the hydrogen gas around PBHs. See section 3.4 and 3.6 in \cite{Ricotti:2007au} in detail. Thus we expect that the constraint on the abundance of PBHs should be tighter if we take into account the formation of binaries. However, the formation of PBH binaries is ignored in \cite{Ricotti:2007au}, and therefore the constraint obtained in this section can be taken as a conservative constraint.



\section{Implications for the event rate of mergers of PBH binaries}
\label{er}

In general PBH binaries could be formed after formation of PBHs. However it is quite difficult to exactly estimate the event rate of mergers of PBH binaries. There are several mechanisms for formation of PBH binaries. See, for example, \cite{Bird:2016dcv,Sasaki:2016jop,Nakamura:1997sm,Ioka:1998nz}. Here we follow the estimation given in \cite{Sasaki:2016jop}.

The physical mean distance between PBHs at matter-radiation equality at the redshift $z=z_{eq}$ is roughly given by
\m
{\bar x}&=&{1\over 1+z_\text{eq}} \({M_\text{pbh}\over  f_\text{pbh}\Omega_\text{CDM}\rho_\text{crit}}\)^{1/3},
\n
where
\m
\rho_\text{crit}&=&{3H_0^2\over 8\pi G}
\n
is the critical density today. The pair of PBHs is supposed to decouple from the expansion of the Universe and forms a gravitational bound system if the average energy density of PBHs over the volume is larger than the background cosmic energy density.
If the motion of the two PBHs is not disturbed, they just coalesce to a single black hole on the free fall time scale. However the tidal force from neighboring black holes provides enough angular momentum to keep the holes from colliding with each other. Taking into account the gravitational waves radiated from the PBH binaries, the event rate (ER) of mergers of PBHs binaries is given by
\begin{equation}
\text{ER(t)}=
\left\{
\begin{aligned}
{f_\text{pbh} \Omega_\text{CDM} \rho_\text{crit}\over M_\text{pbh}}{3\over 58}\[-\({t\over T}\)^{3\over 8}+\({t\over T}\)^{3\over 37}\]{1\over t},\ \ \ \ \ \ \ \ \ \ \ \ \text{for}\ t<t_c,  \\
{f_\text{pbh} \Omega_\text{CDM} \rho_\text{crit}\over M_\text{pbh}} {3\over 58}\({t\over T}\)^{3\over 8}\[-1+\({t\over t_c}\)^{-{29\over 56}} f^{-{29\over 8}}\]{1\over t}, \ \text{for}\ t\geq t_c,
\end{aligned}
\right.
\end{equation}
at the time of $t$, where
\m
t_c&=&Q\bar x^4 f_\text{pbh}^{25/3},\\
T&=& {{\bar x}^4Q/ f_\text{pbh}^4}, \\
Q&=& {3\over 170} \(GM_\text{pbh}\)^{-3}.
\n
Here we focus on the mergers of PBH binaries at low redshift and hence the ER is estimated as ER($t_0$), where $t_0$ is the age of our Universe.

\begin{figure}[]
\begin{center}
\includegraphics[width=5in]{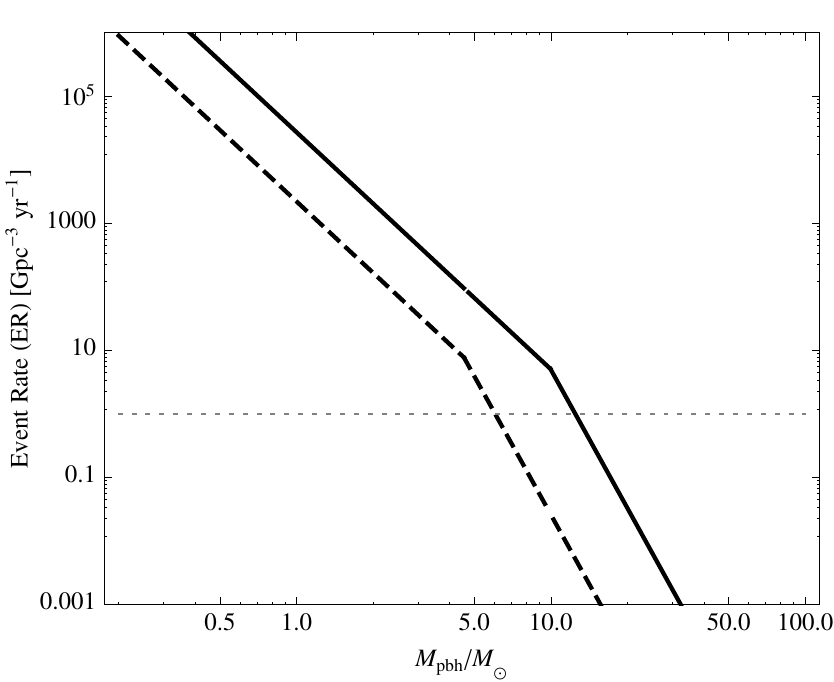}
\end{center}
\caption{The upper limit on the event rate of mergers of PBH binaries. The dashed and solid black lines illustrate the limits at $68\%$ and $95\%$ confidence level respectively. }
\label{fig:ER}
\end{figure}

From Eq.~\eqref{taupbh}, for the PBHs with mass less than $100M_\odot$, the abundance of PBHs with mass $M_\text{pbh}$ is
\m
f_\text{pbh}\simeq 400(\Delta\tau_e)^2 \({M_\text{pbh}\over M_\odot}\)^{-2}.
\n
Considering the constraints on $\Delta\tau_e$ from Planck data obtained in the former section, we illustrate the upper limit on the event rate of mergers of PBH binaries in Fig.~\ref{fig:ER}.
From Fig.~\ref{fig:ER}, we see that the event rate of mergers of PBH binaries with each mass around $30M_\odot$ is less than $0.002$ (Gpc$^{-3}$ yr$^{-1}$) at $95\%$ C.L.. It implies that GW150914 \cite{Abbott:2016blz} is very unlikely produced by a PBH binary. But the event rate of mergers of PBH binaries can be $5$ (Gpc$^{-3}$ yr$^{-1}$) for $M_\text{pbh}=10M_\odot$ and $2000$ (Gpc$^{-3}$ yr$^{-1}$) for $M_\text{pbh}=2M_\odot$.

\section{Summary and discussion}
\label{sd}

In this paper we investigate how the radiations from gas accretion onto PBHs affect the reionization history of our Universe, and furthermore changes the CMB the angular power spectra, including TT, TE, EE and BB. In particular, the polarization power spectra at large scales are changed significantly. Roughly speaking, larger the accretion rate, larger the accretion luminosity. So the CMB data can put a tighter constraint on the large mass PBHs than that on the small mass PBHs. Adopting the Planck data released in 2015, we significantly improve the constraints on the abundance of PBHs in dark matter by around two orders of magnitude. We also apply these new constraints to estimate the event rate of mergers of PBH binaries. Even though GW150914 is unlikely produced by the merger of a PBH binary, we still have opportunity to detect the gravitational waves generated by the solar-mass PBH binaries in the near future.

In addition, because the PBHs disturb the low-$\ell$ CMB polarizations significantly, the improved measurements of CMB polarization at large scales will help us to explore the PBHs. 

In some sense PBHs are a powerful probe of the early universe and could make up a fraction of dark matter in our Universe. Detection of gravitational waves from mergers of PBHs may help us to understand the physics in the early universe in the future.

\vspace{5mm}
\noindent {\bf Acknowledgments}

QGH would like to thank M.~Sasaki for useful conversation.
We acknowledge the use of HPC Cluster of SKLTP/ITP-CAS.
This work is supported by Top-Notch Young Talents Program of China, grants from NSFC (grant NO. 11322545, 11335012, 11575271), the Major Program of the NSFC Grants No. 11690021, and Key Research Program of Frontier Sciences, CAS.




\begin{thebibliography}{99}
\frenchspacing


\bibitem{Hawking:1971ei}
  S.~Hawking,
  Mon.\ Not.\ Roy.\ Astron.\ Soc.\  {\bf 152}, 75 (1971).

\bibitem{Carr:1974nx}
  B.~J.~Carr and S.~W.~Hawking,
  Mon.\ Not.\ Roy.\ Astron.\ Soc.\  {\bf 168}, 399 (1974).

\bibitem{ATLAS:2016hao}
  The ATLAS and CMS Collaborations [ATLAS and CMS Collaborations],
  ATLAS-CONF-2016-036, CMS-PAS-SMP-15-001.


\bibitem{Akerib:2015rjg}
  D.~S.~Akerib {\it et al.} [LUX Collaboration],
  Phys.\ Rev.\ Lett.\  {\bf 116}, no. 16, 161301 (2016)
  doi:10.1103/PhysRevLett.116.161301
  [arXiv:1512.03506 [astro-ph.CO]].

\bibitem{FermiLAT:2011ab}
  M.~Ackermann {\it et al.} [Fermi-LAT Collaboration],
  Phys.\ Rev.\ Lett.\  {\bf 108}, 011103 (2012)
  doi:10.1103/PhysRevLett.108.011103
  [arXiv:1109.0521 [astro-ph.HE]].


\bibitem{Accardo:2014lma}
  L.~Accardo {\it et al.} [AMS Collaboration],
  Phys.\ Rev.\ Lett.\  {\bf 113}, 121101 (2014).
  doi:10.1103/PhysRevLett.113.121101


\bibitem{Abbott:2016blz}
  B.~P.~Abbott {\it et al.} [LIGO Scientific and Virgo Collaborations],
  Phys.\ Rev.\ Lett.\  {\bf 116}, no. 6, 061102 (2016)
  doi:10.1103/PhysRevLett.116.061102
  [arXiv:1602.03837 [gr-qc]].


\bibitem{Abbott:2016nmj}
  B.~P.~Abbott {\it et al.} [LIGO Scientific and Virgo Collaborations],
  Phys.\ Rev.\ Lett.\  {\bf 116} (2016) no.24,  241103
  doi:10.1103/PhysRevLett.116.241103
  [arXiv:1606.04855 [gr-qc]].


\bibitem{Bird:2016dcv}
  S.~Bird, I.~Cholis, J.~B.~Munoz, Y.~Ali-Haimoud, M.~Kamionkowski, E.~D.~Kovetz, A.~Raccanelli and A.~G.~Riess,
  Phys.\ Rev.\ Lett.\  {\bf 116}, no. 20, 201301 (2016)
  doi:10.1103/PhysRevLett.116.201301
  [arXiv:1603.00464 [astro-ph.CO]].


\bibitem{Sasaki:2016jop} 
  M.~Sasaki, T.~Suyama, T.~Tanaka and S.~Yokoyama,
  Phys.\ Rev.\ Lett.\  {\bf 117}, no. 6, 061101 (2016)
  doi:10.1103/PhysRevLett.117.061101
  [arXiv:1603.08338 [astro-ph.CO]].


\bibitem{Kashlinsky:2016sdv}
  A.~Kashlinsky,
  Astrophys.\ J.\  {\bf 823}, no. 2, L25 (2016)
  doi:10.3847/2041-8205/823/2/L25
  [arXiv:1605.04023 [astro-ph.CO]].

\bibitem{Cholis:2016kqi}
  I.~Cholis, E.~D.~Kovetz, Y.~Ali-Haimoud, S.~Bird, M.~Kamionkowski, J.~B.~Munoz and A.~Raccanelli,
  arXiv:1606.07437 [astro-ph.HE].


\bibitem{Carr:2009jm}
  B.~J.~Carr, K.~Kohri, Y.~Sendouda and J.~Yokoyama,
  Phys.\ Rev.\ D {\bf 81}, 104019 (2010)
  doi:10.1103/PhysRevD.81.104019
  [arXiv:0912.5297 [astro-ph.CO]].


\bibitem{Capela:2013yf}
  F.~Capela, M.~Pshirkov and P.~Tinyakov,
  Phys.\ Rev.\ D {\bf 87}, no. 12, 123524 (2013)
  doi:10.1103/PhysRevD.87.123524
  [arXiv:1301.4984 [astro-ph.CO]].


\bibitem{Tisserand:2006zx}
  P.~Tisserand {\it et al.} [EROS-2 Collaboration],
  Astron.\ Astrophys.\  {\bf 469}, 387 (2007)
  doi:10.1051/0004-6361:20066017
  [astro-ph/0607207].

\bibitem{Wyrzykowski:2010mh}
  L.~Wyrzykowski {\it et al.},
  Mon.\ Not.\ Roy.\ Astron.\ Soc.\  {\bf 413}, 493 (2011)
  doi:10.1111/j.1365-2966.2010.18150.x
  [arXiv:1012.1154 [astro-ph.GA]].

\bibitem{Wyrzykowski:2011tr}
  L.~Wyrzykowski {\it et al.},
  Mon.\ Not.\ Roy.\ Astron.\ Soc.\  {\bf 416} (2011) 2949
  doi:10.1111/j.1365-2966.2011.19243.x
  [arXiv:1106.2925 [astro-ph.GA]].


\bibitem{Carr:1981}
  B.~J.~Carr
  MNRAS (1981) 194 (3): 639-668
  doi: 10.1093/mnras/194.3.639


\bibitem{Ricotti:2007au}
  M.~Ricotti, J.~P.~Ostriker and K.~J.~Mack,
  Astrophys.\ J.\  {\bf 680}, 829 (2008)
  doi:10.1086/587831
  [arXiv:0709.0524 [astro-ph]].


\bibitem{Carr:2016drx}
  B.~Carr, F.~Kuhnel and M.~Sandstad,
  arXiv:1607.06077 [astro-ph.CO].



\bibitem{Ade:2015xua}
  P.~A.~R.~Ade {\it et al.} [Planck Collaboration],
  arXiv:1502.01589 [astro-ph.CO].


\bibitem{Douspis:2015nca} 
  M.~Douspis, N.~Aghanim, S.~Ilic and M.~Langer,
  Astron.\ Astrophys.\  {\bf 580}, L4 (2015)
  doi:10.1051/0004-6361/201526543
  [arXiv:1509.02785 [astro-ph.CO]].




\bibitem{Hawking:1974rv}
  S.~W.~Hawking,
  Nature {\bf 248}, 30 (1974).
  doi:10.1038/248030a0

\bibitem{Hawking:1974sw}
  S.~W.~Hawking,
  Commun.\ Math.\ Phys.\  {\bf 43}, 199 (1975)
  Erratum: [Commun.\ Math.\ Phys.\  {\bf 46}, 206 (1976)].
  doi:10.1007/BF02345020


\bibitem{Bondi:1944jm}
  H.~Bondi and F.~Hoyle,
  Mon.\ Not.\ Roy.\ Astron.\ Soc.\  {\bf 104}, 273 (1944).

\bibitem{Bondi:1952ni}
  H.~Bondi,
  Mon.\ Not.\ Roy.\ Astron.\ Soc.\  {\bf 112}, 195 (1952).


\bibitem{Seager:1999bc}
  S.~Seager, D.~D.~Sasselov and D.~Scott,
  Astrophys.\ J.\  {\bf 523}, L1 (1999)
  doi:10.1086/312250
  [astro-ph/9909275].  


\bibitem{Oldengott:2016yjc}
  I.~M.~Oldengott, D.~Boriero and D.~J.~Schwarz,
  arXiv:1605.03928 [astro-ph.CO].  

\bibitem{Lewis:2002ah}
  A.~Lewis and S.~Bridle,
  Phys.\ Rev.\ D {\bf 66}, 103511 (2002)
  [astro-ph/0205436]. 
  
\bibitem{Hayasaki:2009ug} 
  K.~Hayasaki, K.~Takahashi, Y.~Sendouda and S.~Nagataki,
  Publ.\ Astron.\ Soc.\ Jap.\  {\bf 68}, no. 4, 66 (2016)
  doi:10.1093/pasj/psw065
  [arXiv:0909.1738 [astro-ph.CO]].
  

\bibitem{Nakamura:1997sm}
  T.~Nakamura, M.~Sasaki, T.~Tanaka and K.~S.~Thorne,
  Astrophys.\ J.\  {\bf 487}, L139 (1997)
  doi:10.1086/310886
  [astro-ph/9708060].


\bibitem{Ioka:1998nz}
  K.~Ioka, T.~Chiba, T.~Tanaka and T.~Nakamura,
  Phys.\ Rev.\ D {\bf 58}, 063003 (1998)
  doi:10.1103/PhysRevD.58.063003
  [astro-ph/9807018].






\end{thebibliography}
\end{document}